\documentstyle[amssymb,12pt]{article}

\input{tcilatex}

\begin{document}

\title{{\bf {\ Casimir energy for spherical shell in Schwarzschild black hole
background }}}
\author{$^{1,2}$M.R. Setare\thanks{%
E-mail: rezakord@ipm.ir} and  $^{3}$M.B.Altaie\thanks{%
E-mail: maltaie@yu.edu.jo} \\
{$^{1}$Physics Dept. Inst. for Studies in Theo. Physics and
Mathematics(IPM)}\\
{P. O. Box 19395-5531, Tehran, IRAN }
 \\
{$^{2}$Department of Science, Physics group, Kordestan University, Sanandeg,
Iran}\\
\\
{$^{3}$Department of Physics, Yarmouk University, Irbid-Jordan}}
\maketitle

\begin{abstract}
  In this paper, we consider the Casimir energy of massless scalar field which
satisfy Dirichlet boundary condition on a spherical shell. Outside
the shell, the spacetime is assumed to be described by the
Schwarzschild metric, while inside the shell it is taken to be the
flat Minkowski space. Using zeta function regularization and heat
kernel coefficients we isolate the divergent contributions of the
Casimir energy inside and outside the shell, then using the
renormalization procedure of the bag model the divergent parts are
cancelled, finally obtaining a renormalized expression for the
total Casimir energy.

\end{abstract}


\newpage 

\section{Introduction}

The Casimir effect is one of the most interesting manifestations
of nontrivial properties of the vacuum state in quantum field
theory [1,2]. The Casimir effect can be viewed as the polarization
of vacuum by the boundary conditions or geometry. Therefore,
vacuum polarization induced by a gravitational field is also
considered as Casimir effect. Since its first prediction by
Casimir in 1948 \cite{Casimir}, this effect has been investigated
for different fields in different background geometries [4-7].
There is several
  methods for calculating Casimir energy. For instance,  we can mention mode summation,
  Green's function method \cite{mueller}, heat kernel method \cite{{klaus},{bor}}along with appropriate
   regularization schemes such as point separation \cite{chr},\cite{adler}
  dimensional regularization \cite{deser}, zeta function regularization
  \cite{{haw},{Remeo},{Elizalde}}. Recently a general new methods to compute renormalized
   one--loop quantum energies and energy densities are given in
   \cite{{gram1},{gram2}}.
   \newline
It has been shown \cite{{Nugayev1},{Nugayev2}} that particle creation by
black hole in four dimension is a consequence of the Casimir effect for
spherical shell. Also it has been shown that the only existence of the
horizon and of the barrier in the effective potential is sufficient to
compel the black hole to emit black-body radiation with temperature that
exactly coincides with the standard result for Hawking radiation. In \cite
{Nugayev2}, the results for the accelerated-mirror have been used to prove
the above statement. To see more about relation between moving mirrors and
black holes refer to \cite{ful} \newline
Another relation between Casimir effect and Schwarzschild black hole
thermodynamic is the thermodynamic instability. Widom et al \cite
{{Wid},{sass}} showed that the black hole capacity is negative, then an
increase in its energy decreases its temperature. They also showed that the
electrodynamic Casimir effect can also produce thermodynamic instability.%
\newline
The renormalized vacuum expectation value of the stress tensor of the scalar
field in the Schwarzschild spacetime can be obtained by using different
regularization methods.( see Refs. \cite
{{chris},{ant},{bal1},{bal2},{mat1},{mat2}}). $<T_{\nu }^{\mu }>_{ren}$ is
needed , for instance, when we want to study back-reaction, i.e, the
influence that the matter field in a curved background assert on the
background geometry itself. This would be done by solving the Einstein
equations with the expectation value of the energy-momentum tensor as source.%
\newline
Regarding the Nugayev papers \cite{{Nugayev1},{Nugayev2}}, we would like to
investigate the Casimir energy of massless scalar field which is conformally
coupled to the Schwarzschild spacetime and satisfies Dirichlet boundary
condition on a spherical shell. Casimir effect for spherical shells in the
presence of the electromagnetic fields has been calculated several years ago
\cite{{Boyer}, {bal},{Schw}}. The dependence of Casimir energy on the
dimensions of the space for electromagnetic and scalar fields with Dirichlet
boundary conditions in the presence of a spherical shell is discussed in
\cite{{Milton},{Mil}}. The Casimir energy for odd and even space dimensions
and different fields, including the spinor field, and all the possible
boundary conditions have been considered in \cite{Cog}. There it is
explicitly shown that although the Casimir energy for interior and exterior
of a spherical shell are both divergent, irrespective of the number of space
dimensions, the total Casimir energy of the shell remains finite for the
case of odd space dimensions (see also \cite{mil2}). Of some interest are
cases where the field is confined to the inside of a spherical shell. This
is sometimes called the bag boundary condition. The application of Casimir
effect to the bag model is considered for the case of massive scalar field
\cite{bord} and the Dirac field \cite{Eli2}. We will utilize the
renormalization procedure used in the above cases for our problem.\newline
The curvature effects in Schwarzschild background are well studied through
various topics, but the effects of boundaries do not seem to be so generally
familiar. The Casimir energy for the massless scalar fields of two parallel
plates in a two-dimensional Schwarzschild black hole with Dirichlet boundary
conditions has been calculated in Ref.\cite{set1}.\newline
In this paper we would like to investigated the Casimir energy of massless
scalar field for a spherical shell with Dirichlet boundary condition,
Outside the shell, the spacetime is described by the Schwarzschild metric,
while inside the shell is flat Minkowski space. The heat kernel and zeta
function will be utilized to investigate the divergent parts of the vacuum
energy . Heat kernel coefficients and zeta function of the Laplace operator
on a manifold with different boundary conditions, both of them useful tools
to calculate Casimir energies, have been calculated in \cite
{{Birrell},{klaus}}.\newline
The paper is organized as follows: in the second section we briefly review
the Casimir energy inside and outside of spherical shell in terms of zeta
function. Then in section 3 we obtain the heat kernel coefficients for
massless scalar field inside and outside of spherical shell, then we obtain
the divergent part of Casimir energy inside and outside of shell separately.
Section 4 is devoted to the conclusions.

\section{Casimir energy inside and outside of spherical shell}

In what follows as a boundary configuration we shall consider a spherical
shell, outside the shell we consider the spacetime to be described by the
Schwarzschild metric which has the form
\begin{equation}
d^{2}s=-\left( 1-2\frac{m}{r}\right) dt^{2}+\left( 1-2\frac{m}{r}\right)
^{-1}dr^{2}+r^{2}(d\theta ^{2}+\sin ^{2}d\varphi ^{2}),
\end{equation}
while inside the shell the spacetime is the flat Minkowski space. We shall
consider the conformally coupled massless real scalar field $\phi $, which
satisfies
\begin{equation}
\left( \square +\frac{1}{6}R\right) \phi =0,\qquad \square =\frac{1}{\sqrt{-g%
}}\partial _{\mu }\left( \sqrt{-g}g^{\mu \nu }\partial _{\nu }\right) ,
\label{motioneq}
\end{equation}
and propagates inside and outside the shell, $R$ is scalar curvature which
is zero in both the Schwarzschild and Minkowski backgrounds. For the point
on spherical shell the scalar field obeys Dirichlet boundary condition
\begin{equation}
\phi (r=a)=0,  \label{boundcond}
\end{equation}
where $a$ is the radius of spherical shell.\newline
The quantization of the field described by Eq.(2) on the background of
Eq.(1) is standard. Let $\phi _{\alpha }^{(\pm )}(x)$ be complete set of
orthonormalized positive and negative frequency solutions to the field
equation (2), obeying boundary conditions (\ref{boundcond}). The canonical
quantization can be done by expanding the general solution of Eq.(2) in
terms of $\phi _{\alpha }^{(\pm )}$,
\begin{equation}
\phi =\sum_{\alpha }(\phi _{\alpha }^{+}a_{\alpha }+\phi _{\alpha }a_{\alpha
}^{(+)})  \label{expansion}
\end{equation}
and declaring the coefficients $a_{\alpha }$, $a_{\alpha }^{+}$ as operators
satisfying the standard commutation relation for bosonic fields. The vacuum
state $|0>$ is defined as $a_{\alpha }|0>=0$. This state is different from
the vacuum state for black hole geometry without boundaries, $|\bar{0}>$. A
black hole emits particles like a hot body at a  temperature $\frac{\kappa }{%
2\pi }$ where $\kappa $ is a surface gravity of the black hole. Therefore,
we have to considered the Hartle-Hawking state $|\bar{0}>=|H>$, this state
is not empty at infinity, even in the absence of boundary conditions on the
quantum field, but it corresponds to a thermal distribution of quanta at the
Hawking temperature $T=\frac{1}{8\pi m}$. In fact, the state $|H>$ is
related to a black hole in equilibrium with an infinite reservoir of black
body radiation.\newline
The quantum field has a ground state energy
\begin{equation}
E_{0}=\frac{1}{2}\sum_{k}\lambda _{k}^{1/2},
\end{equation}
where the $\lambda _{k}$'s are the one-particle energies with the quantum
number $k$. The vacuum energy is divergent and we shall regularize it by
\begin{equation}
E_{0}=\frac{1}{2}\sum_{k}\lambda _{k}^{1/2-s}\mu ^{2s},\hspace{2cm}Res>2
\end{equation}
where $\mu $ is an arbitrary mass parameter. It is similar to the
subtraction point in the renormalization of perturbative quantum field
theory. After renormalization the ground state energy will become
independent of $\mu $. The one-particle energies are determined by the
eigenvalue equation
\begin{equation}
-\triangle \varphi _{k}=\lambda _{k}\varphi _{k}.
\end{equation}
For the calculations we use the corresponding zeta function
\begin{equation}
\zeta _{A}(s)=\sum_{k}\lambda _{k}^{-s},
\end{equation}
where operator $A$ is given by
\begin{equation}
A=-\square .
\end{equation}
Therefore the regularized vacuum energy inside and outside the spherical
shell are given by
\begin{equation}
E_{reg}^{in}=\frac{1}{2}\zeta _{A}^{in}(s-1/2)\mu ^{2s},\hspace{1cm}%
E_{reg}^{out}=\frac{1}{2}\zeta _{A}^{out}(s-1/2)\mu ^{2s}.
\end{equation}

\section{Zeta function and Heat-Kernel coefficients}

The general structure of the ultraviolet divergencies can be obtained from
the heat kernel expansion. For this reason one can represent the zeta
function in Eq.(8) by an integral
\begin{equation}
\zeta _{A}(s)=\frac{1}{\Gamma (s)}\int_{0}^{\infty }dtt^{s-1}K(t),
\end{equation}
where
\begin{equation}
K(t)=(4\pi t)^{-3/2}\sum_{k}\exp (-\lambda _{k}t),
\end{equation}
is the heat kernel. Now the ultraviolet divergencies of the vacuum energy
are determined from the behaviour of the integrand in Eq.(11) at the lower
integration limit and , hence, from the asymptotic expansion of the heat
kernel for $t\rightarrow 0$
\begin{equation}
K(t)\sim \frac{1}{(4\pi t)^{3/2}}\sum_{k=0,1/2,1,...}B_{k}t^{k}.
\end{equation}
This expansion is known for a very general manifold, if the underlying
manifold is without boundary, only coefficients with integer numbers enter,
otherwise half integer powers of $t$ are present. The $B_{k}$ are given by
\begin{equation}
B_{k}=\int_{M}dva_{k}(x)+\int_{S^{2}}dsc_{k}(y),
\end{equation}
the Seely-de Witt coefficients $a_{k}(x)$ vanish for half-odd integers,
these coefficients are independent of the applied boundary condition, but
the coefficients do depend on the spin of the field in question \cite
{{Birrell},{bl}}. The coefficients $c_{k}$ are functions of the second
fundamental form of the boundary (extrinsic curvature), the induced geometry
on the boundary (intrinsic curvature), and the nature of boundary conditions
imposed. The simplest first of $a_{k}$ and $c_{k}$ coefficients for a
manifold with boundary are given in \cite{Birrell}
\begin{equation}
a_{0}(x)=1,
\end{equation}
\begin{equation}
a_{1}(x)=(\frac{1}{6}-\xi )R,
\end{equation}
\begin{equation}
a_{2}(x)=\frac{1}{180}R_{\alpha \beta \gamma \delta }R^{\alpha \beta \gamma
\delta }-\frac{1}{180}R^{\alpha \beta }R_{\alpha \beta }-\frac{1}{6}(1/5-\xi
)\Box R+\frac{1}{2}(1/6-\xi )^{2}R^{2},
\end{equation}
where $\xi $ is the coupling constant between the scalar field and the
gravitational field, for conformally coupling $\xi =1/6$, $R_{\alpha \beta
\gamma \delta }$, $R_{\alpha \beta }$ and $R$ are respectively, Riemann,
Ricci and scalar curvature tensors. The $c_{k}$ coefficients for Dirichlet
boundary condition are as follow \cite{Birrell}
\begin{equation}
c_{0}=0,
\end{equation}
\begin{equation}
c_{1/2}=-\frac{\sqrt{\pi }}{2},
\end{equation}
\begin{equation}
c_{1}=\frac{1}{3}K-\frac{1}{2}f^{(1)},
\end{equation}
\begin{equation}
c_{3/2}=\frac{\sqrt{\pi }}{2}((\frac{-1}{6}\hat{R}-\frac{1}{4}%
R_{ik}N^{i}N^{k}+\frac{3}{32}(trK)^{2}-\frac{1}{16}trK^{2})+\frac{5}{16}%
trKf^{(1)}-\frac{1}{4}f^{(2)}),
\end{equation}
\begin{eqnarray}
c_{2} &=&\frac{1}{3}(\frac{1}{6}-\xi )R(trK)+\frac{1}{3}(\frac{3}{20}-\xi
)\nabla _{l}N^{l}-\frac{1}{90}R_{lk}N^{l}N^{k}(trK)+\frac{1}{30}%
R_{iljk}N^{l}N^{k}K^{ij} \\
&-&\frac{1}{90}R_{il}K^{il}+\frac{1}{315}[\frac{5}{3}%
(trK)^{3}-11(trK)(trK^{2})+\frac{40}{3}(trK^{3})]+\frac{1}{15}\Box (trK)
\nonumber
\end{eqnarray}
where $f^{(i)}$ are the $i$'th normal derivative of the function $f$, $%
R_{iljk}$ $R_{ik}$and $\hat{R}$ are respectively Riemann, Ricci and the
scalar curvature on the boundary, $K$ is extrinsic curvature tensor on the
boundary
\begin{equation}
K_{ij}=\nabla _{i}N_{j},
\end{equation}
where $N_{j}$ is unit normal vector.\newline
Given the expression in Eq.(11), it is easy to isolate the pole in $\zeta
_{A}(s)$ since
\begin{equation}
\zeta _{A}(s)=\frac{1}{\Gamma (s)}\int_{0}^{1}dtt^{s-1}K(t)+\frac{1}{\Gamma
(s)}\int_{1}^{\infty }dtt^{s-1}K(t)
\end{equation}
Due to the exponential fall of $K(t)$ for large $t$, it is clear that the
second term in the above expression is perfectly finite function of complex $%
s$. Observe that asymptotic expansion Eq.(11) implies that $\zeta _{A}(s)$
has a pole structure given by
\begin{equation}
\zeta _{A}(s)=\frac{1}{(4\pi )^{3/2}\Gamma (s)}\sum_{k=0,1/2,1,...}\frac{%
B_{k}}{k+s-3/2}+finite
\end{equation}
Thus $\zeta _{A}(s)$ has a simple pole whenever $s=3/2-k$, expect at $s=0$
where any possible pole is cancelled by that of $\Gamma (s)$. The residue of
the pole is given by
\begin{equation}
\frac{B_{3/2-s}}{(4\pi )^{3/2}\Gamma (s)}.
\end{equation}
However, $\zeta _{A}(s)$ is analytic at $s=0$, and one can calculate simply
the values of $\zeta -$function and its derivative at this point. Now, in
order to determine the Casimir energy inside and outside the spherical
shell, we must set $s=-1/2$ and we have a pole with nonzero residue if $%
B_{2}\neq 0$. Then for the case of a massless free scalar field, the only
remaining contributions are from $B_{2}$. These contribution for inside and
outside the shell are divergent. Considering only the inner space,
divergence appear and it is necessary to introduce contact term and perform
a renormalization of its coupling. Result for massive scalar field contain
new ultraviolet divergent terms in addition to that occurring in the
massless case as has been discussed in \cite{bord}. However, when we
consider both region of space, for free massless scalar field in flat space
divergent part inside and outside the shell cancel out each other, then we
do not need to introduce contact term, but for massless scalar field in
curved space, similar to free massive case, when we add the interior and
exterior energies to each other, there will be contributions which are
divergent \cite{{set2},{set3}}. For the case of a massless scalar field in
curved space the divergent part of vacuum energy in zeta function
regularization is proportional with $B_{2}^{tot}$ which is
\begin{equation}
B_{2}^{tot}=B_{2}^{in}+B_{2}^{out}.
\end{equation}
Now using Eqs.(10) and (25) we can write
\begin{equation}
E_{div}^{in}=\frac{\mu ^{2s}}{2(4\pi )^{3/2}\Gamma (s-1/2)}B_{2}^{in},
\end{equation}
Similarly for outside region
\begin{equation}
E_{div}^{out}=\frac{\mu ^{2s}}{2(4\pi )^{3/2}\Gamma (s-1/2)}B_{2}^{out}.
\end{equation}
Since the inside region is assumed to be flat and since the outside space is
considered to be a Schwarzschild background, therefore
\begin{equation}
a_{2}^{in}(x)=0,
\end{equation}
and

\begin{equation}
a_{2}^{out}(x)=\frac{1}{180}R_{\alpha \beta \gamma \delta }R^{\alpha \beta
\gamma \delta }.
\end{equation}
The coefficients $c_{2}$ contains only odd powers of the second fundamental
form $K$, if we consider infinitely thin boundary, which means that boundary
consist of two oppositely oriented faces separated by an infinitesimal
distance, then the second fundamental forms are equal and opposite on the
two face of the boundary, and consequently we have \cite{bl}
\begin{equation}
c_{2}^{in}+c_{2}^{out}=0.
\end{equation}
Therefore
\begin{equation}
B_{2}^{tot}=B_{2}^{in}+B_{2}^{out}=\int_{M}a_{2}^{out}(x)dv.
\end{equation}
Then the total divergent energy is given by
\begin{equation}
E_{div}^{tot}=\frac{\mu ^{2s}}{2(4\pi )^{3/2}\Gamma (s-1/2)}%
(B_{2}^{in}+B_{2}^{out})=\frac{\mu ^{2s}}{2(4\pi )^{3/2}\Gamma (s-1/2)}%
\int_{M}a_{2}^{out}(x)dv,
\end{equation}
Therefore the Casimir energy for this general case becomes divergent. At
this stage we recall that $E_{0},$ as given by Eq.(6), is only one part of
total energy. There is also a classical part. The total energy of the shell
maybe written as
\begin{equation}
E^{tot}=E_{0}+E_{class}
\end{equation}
We can try to absorb $E_{div}$ into the classical energy. This technique of
absorbing an infinite quantity into a renormalized physical quantity is
familiar in quantum field theory and quantum field theory in curved space
\cite{Birrell}. Here, we use a procedure similar to that of bag model \cite
{{bord},{Eli2}}, there is some history of such notions going back to Milton
paper \cite{mil2}, (to see application of this renormalization procedure in
Casimir effect problem in curved space refer to \cite
{{set2},{set3},{dme},{kh}}). The classical energy of spherical shell may be
written as,
\begin{equation}
E_{class}=Pa^{3}+\sigma a^{2}+Fa+K+\frac{h}{a},
\end{equation}
where $P$ is pressure, $\sigma $ is surface tension and $F$, $K$,$h$ do not
have special names. The classical energy is expressed in a general
dimensionally suitable form which depends on power of $a$, this definition
is useful for its renormalization. In order to obtain a well defined result
for the total energy, we have to renormalize only pressure of classical
energy according to the below:
\begin{equation}
P\rightarrow P-\frac{\mu ^{2s}}{2(4\pi )^{3/2}a^{3}\Gamma (s-1/2)}%
\int_{M}a_{2}^{out}(x)dv.
\end{equation}
According to the renormalization procedure, we have to subtract from
regularized expression for vacuum energy Eq.(10) the above divergent term .
After subtracting this contribution from $E_{0}$ we denote it by
\begin{equation}
E_{0}^{ren}=E_{0}-E_{div}^{tot},
\end{equation}
where
\begin{equation}
E_{0}^{ren}=E_{0}^{(in)ren}+E_{0}^{(out)ren}
\end{equation}
The renormalized Casimir energy inside of the spherical shell for massless
free scalar field with Dirichlet boundary conditions in flat Minkowski space
is given by \cite{Cog}
\begin{equation}
E_{0}^{(in)ren}=\frac{0.008873}{2a}.
\end{equation}
But for outside the shell in our problem we have
\begin{equation}
E_{0}^{(out)ren}=E_{0}^{out}-E_{div}^{out}=\frac{1}{2}\zeta
_{A}^{out}(s-1/2)\mu ^{2s}-\frac{\mu ^{2s}}{2(4\pi )^{3/2}\Gamma (s-1/2)}%
\int_{M}a_{2}^{out}(x)dv
\end{equation}
Therefore we can write the renormalized vacuum energy for the considered
system as
\begin{equation}
E_{0}^{ren}=\frac{0.008873}{2a}+\frac{1}{2}\zeta _{A}^{out}(s-1/2)\mu ^{2s}-%
\frac{\mu ^{2s}}{2(4\pi )^{3/2}\Gamma (s-1/2)}\int_{M}a_{2}^{out}(x)dv.
\end{equation}

\section{conclusions}

In this paper we have developed a systematic approach to the
calculation of the Casimir energy of a massless scalar field in
the presence of a spherical shell as a boundary configuration. The
spacetime outside the shell is described by the Schwarzschild
metric, while inside the shell it is the flat Minkowski space. For
the point on the spherical shell, the scalar field obeys Dirichlet
boundary condition.\\
The renormalized vacuum expectation value of the stress tensor of
the scalar field in the curved spacetime , is needed  for
instance, when we want to study back-reaction, i.e, the influence
that the matter field in a curved background assert on the
background geometry itself. It has been shown
  \cite{{Nugayev1},{Nugayev2}} that particle creation by black
  hole in four dimension is as a consequence of the Casimir effect
  for spherical shell. It has been shown that the only existence
  of the horizon and of the barrier in the effective potential is
  sufficient to compel the black hole to emit black-body radiation with
  temperature that exactly coincides with the standard result for
  Hawking radiation. In \cite{Nugayev2}, the results
  for the accelerated-mirror have been used to prove above
  statement. Regarding the Nugayev papers \cite{{Nugayev1},{Nugayev2}}, we
  have investigated the Casimir energy of massless scalar field which is
conformally coupled to the Schwarzschild spacetime and satisfies
Dirichlet boundary condition on a spherical shell.
\\ Using zeta function regularization and heat
kernel coefficients we obtain the divergent contributions for the
Casimir energy inside and outside the shell . When we consider
both region of space, for free massless scalar field in flat
space, the divergent parts inside and outside cancel out each
other, then we do not to introduce a contact term, but for
massless scalar field in curved space, similar to free massive
case, when we add the interior and exterior energies to each
other, there are contributions which are divergent. For a massless
scalar field the divergent part of vacuum energy in zeta function
regularization is proportional to $B_{2}^{tot}$, then the
renormalization procedure become necessary in this situation. For
this purpose one must introduce the classical energy and try to
absorb the divergent part into it. In this paper we used a
procedure similar to that of bag model \cite{{bord},{Eli2}} for
renormalization, according to which we have to subtract from
regularized expression for vacuum energy in Eq.(10) the divergent
term, consequently we obtained the renormalized vacuum energy for
considered system given by Eq.(42).

\vspace{3mm}

{\small . }

\end{document}